\documentclass[final,5p,times,twocolumn]{elsarticle}

\usepackage{graphicx}
\usepackage{amssymb}
\journal{Nuclear Instruments and Methods A}

\begin{document}

\begin{frontmatter}



\title{Beam Studies of the Segmented Resistive WELL: a Potential Thin Sampling Element for Digital Hadron Calorimetry \\ Presented at the $13^{th}$ Vienna Conference on Instrumentation, February 2013}

% if there is only one institution, use this form:
%\author{John Author, Giovanna Autore}
%\address{University of Wisdom, Physics City, Scienceland}

% else, use optional labels to link authors explicitly to addresses,
% as shown below:
\author[A]{Lior Arazi} 
\author[B]{Carlos Davide Rocha Azevedo}
\author[A]{Amos Breskin}
\author[A]{Shikma Bressler}
\author[A]{Luca Moleri}
\author[C]{Hugo Natal da Luz}
\author[D]{Eraldo Oliveri}
\author[A]{Michael Pitt}
\author[A]{Adam Rubin}
\author[C]{Joaquim Marques Ferreira dos Santos}
\author[B]{Jo\~ao Filipe Calapez Albuquerque Veloso}
\author[E]{Andrew Paul White}

\address[A]{Weizmann Institute of Science, Rehovot, Israel}
\address[B]{I3N - Physics Department, University of Aveiro, 3810-193 Aveiro, Portugal}
\address[C]{University of Coimbra, Coimbra, Portugal}
\address[D]{CERN, Geneva, Switzerland}
\address[E]{University of Texas, Arlington, USA}

\begin{abstract}

Thick Gas Electron Multipliers (THGEMs) have the potential of constituting thin, robust sampling elements in Digital Hadron Calorimetry (DHCAL) in future colliders. We report on recent beam studies of new single- and double-THGEM-like structures; the multiplier is a Segmented Resistive WELL (SRWELL) - a single-faced THGEM in contact with a segmented resistive layer inductively coupled to readout pads. Several 10$\times$10 cm$^2$ configurations with a total thickness of 5-6 mm (excluding electronics) with 1 cm$^2$ pads coupled to APV-SRS readout were investigated with muons and pions. Detection efficiencies in the 98$\%$ range were recorded with average pad-multiplicity of $\sim$1.1. The resistive anode resulted in efficient discharge damping, with potential drops of a few volts; discharge probabilities were $\sim$10$^{-7}$ for muons and $\sim$10$^{-6}$ for pions in the double-stage configuration, at rates of a few kHz/cm$^2$. Further optimization work and research on larger detectors are underway.

\end{abstract}

\begin{keyword}
Micropattern gaseous detectors (MPGD) \sep THGEM \sep SRWELL \sep Digital hadron calorimetry (DHCAL) \sep Resistive electrode \sep SRS \sep ILC \sep CLIC
%Latex template file
%
%%% PACS codes here, in the form: \PACS code \sep code
%%% Find PACS codes here: http://www.aip.org/pacs/pacs2010/individuals/pacs2010_regular_edition/index.html
%
%%% MSC codes here, in the form: \MSC code \sep code
%%% or \MSC[2008] code \sep code (2000 is the default)
%
\end{keyword}
\end{frontmatter}
%
%%% \linenumbers
%
%%% main text
\section{Introduction}
The Thick Gas Electron Multiplier (THGEM) [1] is a simple and robust electrode suitable for large area detectors, which can be economically produced by industrial Printed Circuit Board (PCB) methods. Its properties and potential applications are reviewed in [2,3]; recent progress can be found in [4-7]. One possible application of THGEM-based detectors is in Digital Hadronic Calorimeters (DHCAL), of the kind proposed for the ILC/CLIC-SiD experiment [8,9]. In this project, the calorimeter design dictates very narrow sampling elements, in the sub-centimeter range, with a lateral pixel size of 1$\times$1 cm$^2$. Additional requirements are high detection efficiency ($>$95$\%$) and minimum pad multiplicity (number of pads activated per particle).

RPCs presently constitute the baseline technology for the SiD DHCAL, with 94$\%$ efficiency and pad multiplicity of 1.6 [10]; other solutions have been investigated, e.g. MICROMEGAS with 98$\%$ efficiency and multiplicity of 1.1 [11] and Double-GEMs, with 95$\%$ efficiency and pad multiplicity of 1.3 [12] (all results are for muons).

Recently, THGEM-based sampling elements were proposed; they were investigated with muons and pions, primarily in single- and double-THGEM configurations with direct charge collection on readout pads, separated by a 2 mm induction gap from the multiplier [13]. The potential value of this concept for DHCAL was demonstrated, leaving room for further optimization, in terms of stability in hadronic beams, efficiency, multiplicity and overall thickness.

We report here on the results of our latest beam study, conducted at the CERN SPS/H4 RD51 beam-line with 150 GeV/c muons and pions; further substantial progress was made with a novel THGEM-like configuration, the Segmented Resistive WELL (SRWELL). More details can be found elsewhere [14].

\section{Experimental setup and methodology}
The SRWELL, first suggested in [13], is shown schematically in figures 1 and 2; it is a THGEM that is copper-clad on its top side only, whose bottom is closed by a resistive anode. The anode consists of a 0.1 mm thick FR4 sheet patterned with a square grid of narrow copper lines, with the entire area coated with a resistive film (e.g. graphite mixed with epoxy [15]). Avalanche-induced signals are recorded inductively on a pad array located below the FR4 sheet. The grid lines on the resistive anode correspond to the inter-pad boundaries; they serve to prevent charge spreading across neighboring pads by allowing for rapid draining of the avalanche electrons diffusing across the resistive layer. The resistive layer itself (of $\sim$10-20 M$\Omega$/square) serves to significantly reduce the energy of occasional discharges. The closed-bottom geometry, similar in its field shape to the WELL [16] and C.A.T. [17], reduces the total thickness of the detector; it also results in attaining higher gain at a given applied voltage, compared to a standard THGEM with an induction gap [13]. The SRWELL has a segmented square hole-pattern with ``blind" copper strips above the grid lines; these prevent more energetic discharges in holes located above the metal grid.   

\begin{figure}[hbt] 
\centering 
\includegraphics[width=\columnwidth,keepaspectratio]{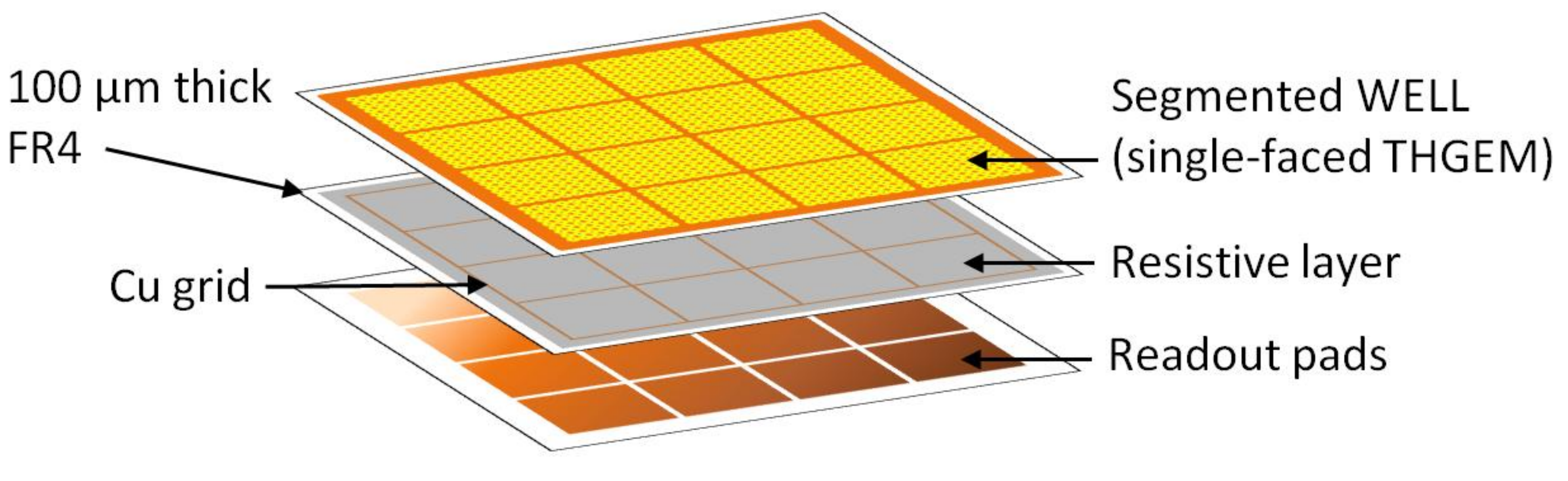}
\caption{The three layers comprising the SRWELL. Bottom: readout pad array (here 4$\times$4); middle: resistive layer on top of a copper grid (on FR4 sheet); top: segmented single-faced THGEM. The layers are assembled one on top of the other in direct contact (see Fig. 2).}
\label{fig:SRWELL_layers}
\end{figure}

Two basic detector configurations were investigated (figure 2): one comprising a single-stage SRWELL and the other a double-stage structure with a standard THGEM followed by an SRWELL. In both cases the electrodes were 10$\times$10 cm$^2$ in size. Based on previous experience with neon-based gas mixtures, which allow for high-gain operation at relatively low voltages [4], the detectors were operated in Ne/5$\%$CH$_4$ at 1 atm, in a typical flow of a few l/h; a minimally ionizing particle (MIP) passing through this gas mixture generates, on the average, $\sim$60 electron-ion pairs per cm along its track [18].

In the single-stage detector the SRWELL was either 0.4 or 0.8 mm thick, with corresponding drift gaps of 5.5 and 5 mm. In the double-stage configuration, both the THGEM and SRWELL were 0.4 mm thick; the transfer gap between them was 1.5 mm wide and the drift gap was 2.5, 3 or 4 mm wide. The total thickness of the detector from the resistive anode to the drift electrode was thus between 4.8 and 6.3 mm. The THGEM and SRWELL electrodes were manufactured by Print Electronics Israel [19] by mechanical drilling of 0.5 mm holes in FR4 plates, Cu-clad on one or two sides, followed by chemical etching of 0.1 mm wide rims around each hole. In the double-stage detectors the THGEM had an hexagonal hole pattern with a pitch of 1 mm; the SRWELL square hole pattern had a pitch of 0.96 mm, with 0.86 mm wide ``blind" strips above the grid lines (1.36 mm between the centers of holes on the opposite sides of the strip). The resistive layers had a surface resistivity of 10-20 M$\Omega$/square; the FR4 sheet serving as the base of the resistive anode was 0.1 mm thick. The grid patterned on the FR4 sheet comprised 0.1 mm wide copper lines, defining an array of 8$\times$8 squares, 1 cm$^2$ each, matching the 8$\times$8 readout pad array patterned here on a 3.2 mm thick FR4 plate located below the anode.

\begin{figure}[hbt] 
\centering 
\includegraphics[width=\columnwidth,keepaspectratio]{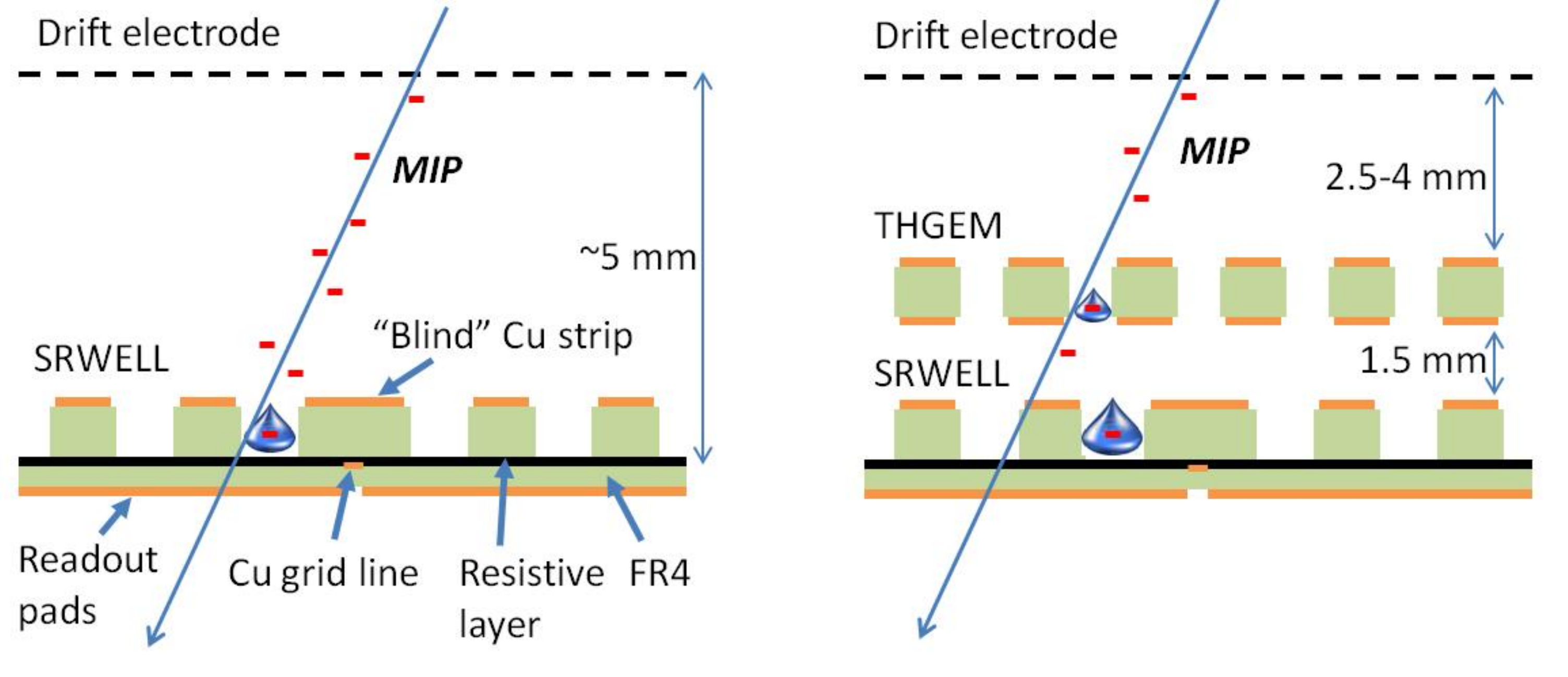}
\caption{The two detector configurations investigated in this work. Left: single-stage SRWELL; right: double-stage detector with a standard THGEM multiplier followed by an SRWELL.}
\label{fig:single_and_double_schemes}
\end{figure}

Data acquisition was done using the new CERN-RD51-SRS readout electronics [20], with the 64-pad array read by a single SRS analog 128-channel APV25 chip [21]. External triggering and tracking were done using the RD51 tracker/telescope setup [22], comprising three 10$\times$10 cm$^2$ scintillators in coincidence with three MICROMEGAS tracking units, each equipped with two APV25 chips. The three tracker detectors and the investigated detector shared the same external trigger and front-end card (FEC), enabling event-by-event matching and track reconstruction. This permitted measuring both the global average values of the detection efficiency and pad multiplicity and their local, position-dependent values (e.g. its variation at the pad boundaries). The low-noise electronics enabled operating the detectors at relatively modest gas gains of $\sim$2000-3000. 

The detector electrodes were biased individually through a CAEN SY2527 HV system. The voltage and current on each HV channel were monitored and stored using the RD51 slow-control system [23], allowing for measuring the rate and magnitude of occasional discharges (e.g. momentary voltage drops, accompanied by current pulses).

The detectors were investigated in a broad low-rate (10-20 Hz/cm$^2$) muon beam, and in narrow, $\sim$1cm$^2$, pion beams; the pion rates were varied between $\sim$0.5 kHz/cm$^2$ to $\sim$70 kHz/cm$^2$, with the majority of the data taken at rates of up to a few kHz/cm$^2$.

Average and local values of the detector efficiency and pad multiplicity were studied using selected tracker events. Pads were considered as activated if their signal was above a pad-specific threshold (set according to its noise level). The detector efficiency was defined as the fraction of tracks where a corresponding cluster of pads was found with its calculated center of gravity not more than 10 mm away from the track projection on the detector. These same tracks were used to calculate the average pad multiplicity by counting the number of pads activated per event. For more details see [14].

\section{Results}

Studies on single-stage detectors included two configurations: one with a 0.4 mm thick SRWELL and a 5.5 mm drift gap and the other with a 0.8 mm thick SRWELL and a 5 mm drift gap. In a muon beam, the former reached 97$\%$ global efficiency at an average pad multiplicity of 1.2, and the latter (0.8 mm thick SRWELL) displayed 98$\%$ global efficiency already at 1.1 multiplicity. The measured Landau pulse-height distributions were well above noise at gains of $\sim$1500-2000. Discharge probabilities with muons were of the order of 10$^{-6}$ for both configurations. However, with pions both configurations displayed a gain drop by a factor of $\sim$2 at the above operating conditions, with a $\sim$5-10 fold increase in discharge probability; this resulted in lower detection efficiencies with pions in both cases. The observed discharges could be divided, in both detector configurations, into two distinct groups: (a) a vast majority of micro-discharges, involving small ($\sim$10-15 V) voltage drops with a typical recovery time of $\sim$2 seconds; (b) a small fraction of discharges involving large voltage drops ($\sim$100-200 V), with longer recovery times (a few seconds, depending on the size of the voltage drop). Of the two detectors, the 0.8 mm thick SRWELL appeared to be more stable, but this requires further study and more precise quantification.  

Studies of the double-stage detectors were done with 0.4 mm thick THGEM and SRWELL electrodes. The transfer gap was kept at 1.5 mm and the drift gap was varied between 2.5 and 4 mm. The efficiencies recorded with muons were similar to those obtained with the single-stage detectors, albeit shifted to slightly higher multiplicities. For example, the 4 mm drift, double-stage detector reached 97$\%$ global efficiency at an average multiplicity of 1.15; the 3 mm drift, double-stage detector reached 94$\%$ efficiency at a multiplicity of 1.2. Discharge probabilities with muons were extremely low, of the order of 10$^{-7}$, for the 4 mm drift double-stage. Figure 3 shows the global efficiency vs. average pad multiplicity for the 0.8 mm thick single-stage SRWELL and the double-stage THGEM/SRWELL with 4 mm drift. Measurement details are provided in [14].

\begin{figure}[hbt] 
\centering 
\includegraphics[width=\columnwidth,keepaspectratio]{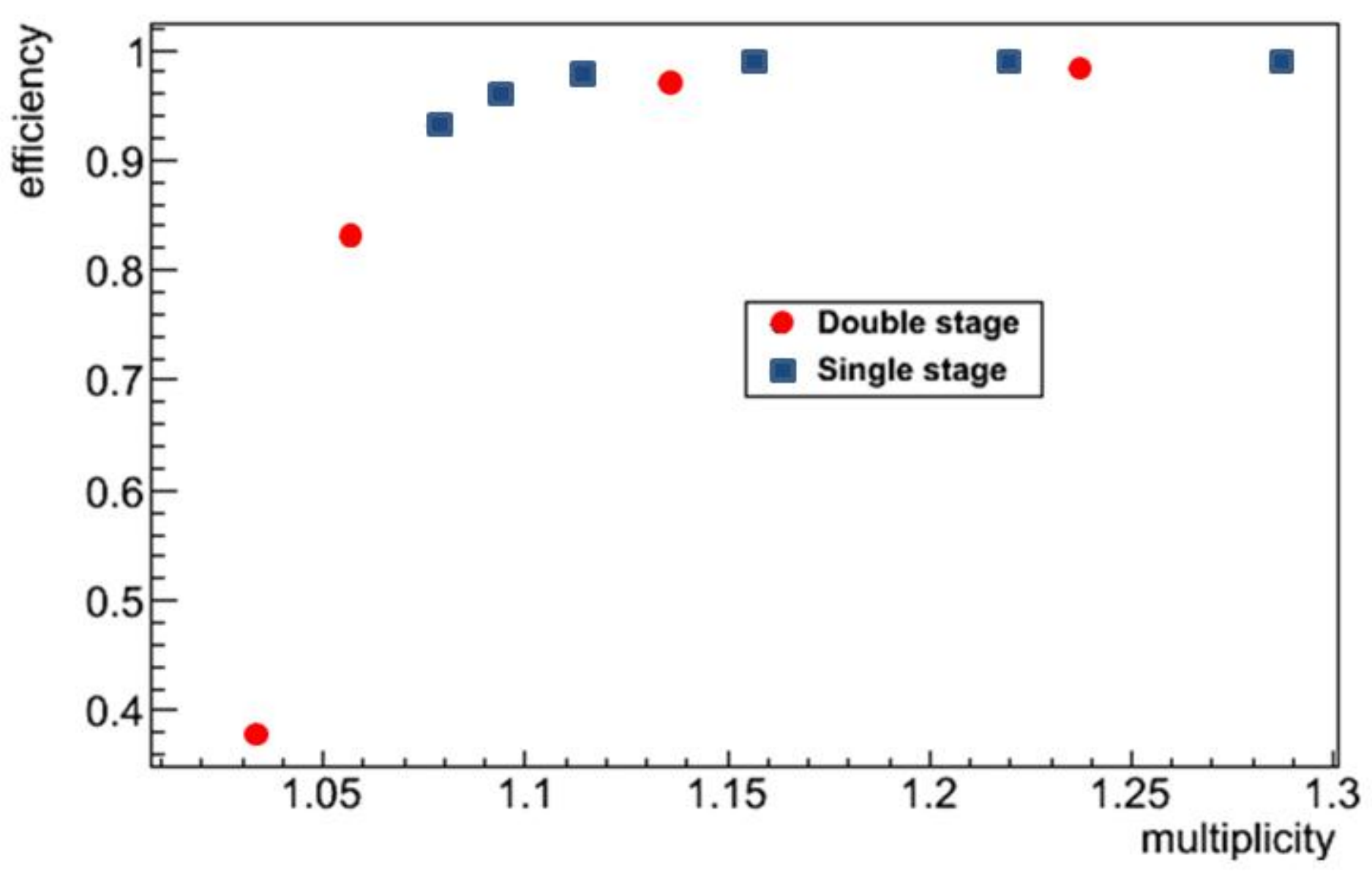}
\caption{Global detection efficiency vs. average pad multiplicity of the 0.8 mm thick single-stage SRWELL with 5 mm drift gap and the double-stage THGEM/SRWELL with 4 mm and 1.5 mm drift and transfer gaps.}
\label{fig:efficiency_vs_multiplicity}
\end{figure}

Unlike the single-stage detectors, no gain drop was observed when switching from muons to pions in the double-stage detectors; figure 4 compares the pulse-height distributions measured for both particles with the double-stage detector of a 4 mm drift gap, under the same operation voltages. Although occasional discharges occurred with pions for the double-stage 4 mm drift detector, their probability, at rates of a few kHz/cm$^2$, was of the order of 10$^{-6}$, and the voltage drops (on the SRWELL top) were all minute, limited to $\sim$3 V, with a recovery time of $\sim$1 second (no large discharges were observed). The efficiency for pions was similar to that obtained with muons ($\sim$95$\%$ and above). A comparison between runs with and without these micro-discharges showed that their effect is negligible in terms of the detection efficiency. Moreover, the presence of micro-discharges had no effect on the data acquisition system, which operated stably even in high rate pion beams.

\begin{figure}[hbt] 
\centering 
\includegraphics[width=\columnwidth,keepaspectratio]{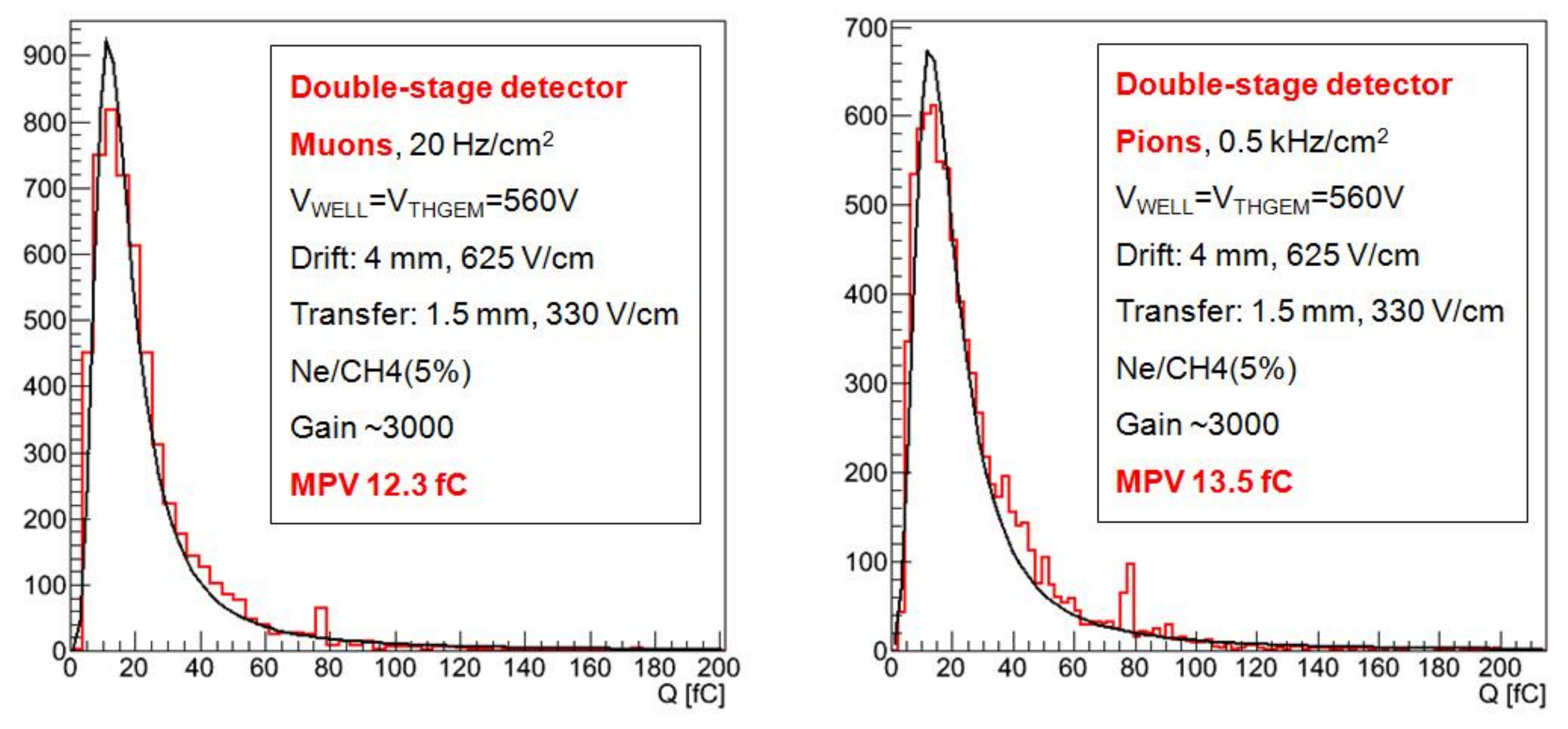}
\caption{Pulse height (Landau) distributions for muons (left) and pions (right) measured for the double-stage detector of 4 mm drift gap. The parameters and operation conditions are given in the figures. No gain drop was observed with pions in this configuration.}
\label{fig:Landaus_double}
\end{figure}

The ability to accurately match events between the tracker and investigated detectors permitted studying the dependence of the local efficiency and pad multiplicity on the track position relative to the pad boundary. The results are shown in figure 5; essentially no drop in local efficiency occurred above the ``blind" SRWELL strips in both configurations; the local increase in pad multiplicity above the inter-pad boundary resulted from charge sharing between holes on the opposite sides of the copper strip (see Fig. 1).

\begin{figure}[hbt] 
\centering 
\includegraphics[width=\columnwidth,keepaspectratio]{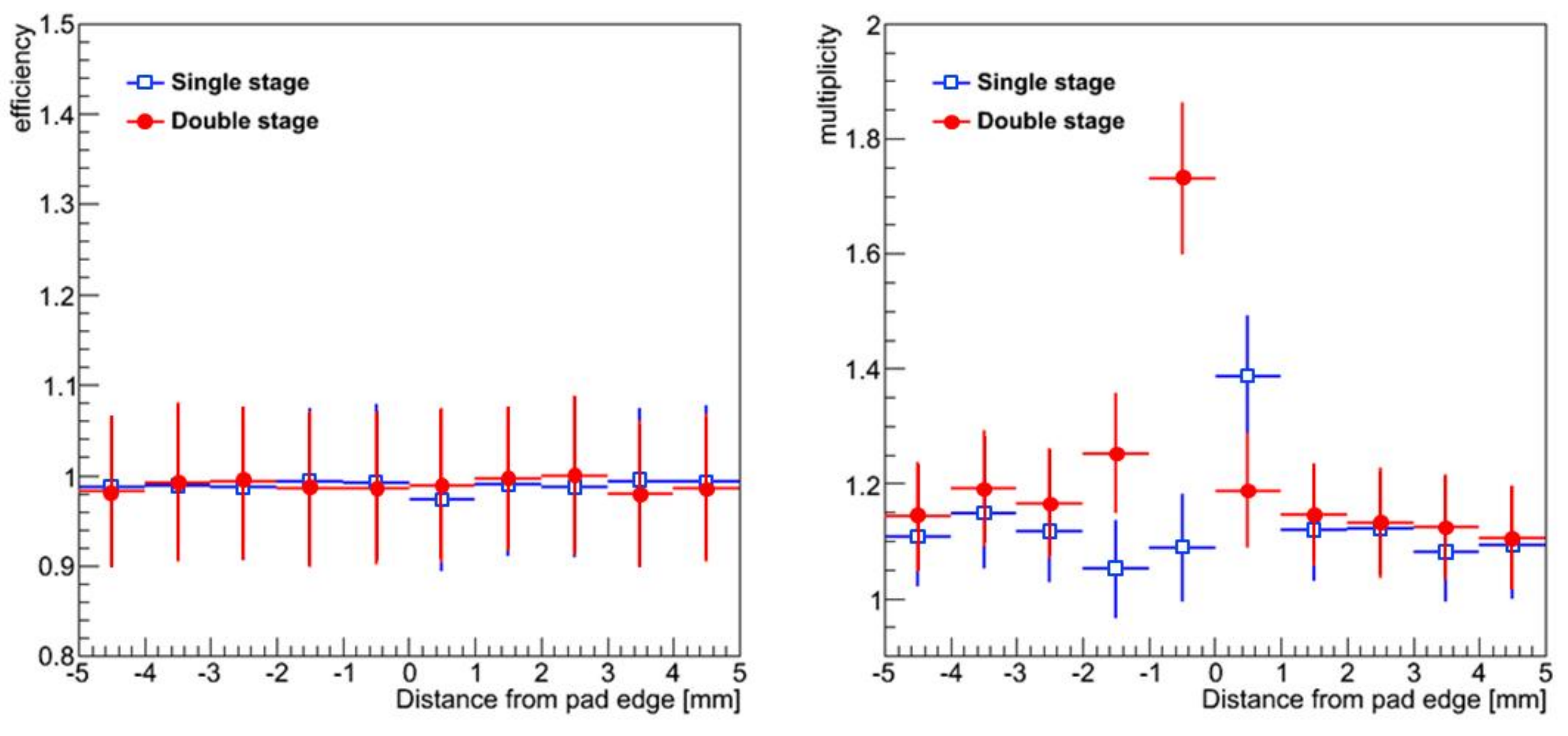}
\caption{Local detection efficiency (left) and pad multiplicity (right) as a function of the muon-hit distance from the pad edge for the single-stage 0.8 SRWELL and the 4 mm drift double-stage detectors.}
\label{fig:Local_efficiency_and_multiplicity}
\end{figure}

\section{Summary and discussion}
The beam tests described in this work investigated, for the first time, structures based on the Segmented Resistive WELL (SRWELL) concept. This new THGEM-variant has several key advantages which make it a promising candidate for Digital Hadronic Calorimetry: (1) By removing the ``standard" induction gap, it allows for a significant reduction in width - a critical feature in applications such as the SiD experiment; the total thickness of the detector configurations studied in this test (excluding readout electronics) was 5-6 mm; (2) The resistive anode effectively quenches occasional discharges, whose magnitude, in the double-stage configuration, was limited to $\sim$3 V with $\sim$1 s recovery time with no effect on the detection efficiency or stability of the electronic readout system; (3) The copper grid underneath the resistive layer significantly reduces the cross-talk between neighboring pads, limiting the multiplicity to $\sim$1.1-1.15; the latter is mostly due to particles inducing avalanches on more than one hole; (4) The detection efficiency with muons is exceptionally high: 98$\%$ at a multiplicity of 1.1 for the single-stage 0.8 mm SRWELL and 97$\%$ at a multiplicity of 1.15 for the 4 mm drift double-stage THGEM/SRWELL. Finally - the SRWELL, like the standard THGEM, is a robust electrode which is essentially immune to spark damage and which can be readily and economically produced over large areas by industrial methods. The combination of the above properties make the SRWELL-based detectors highly competitive compared to the other technologies considered for the SiD-DHCAL.

Single-stage detectors are obviously advantageous in terms of cost when considering large-area applications such as the DHCAL. While their efficiency and multiplicity figures with muons are very convincing, the gain drop in the single-stage SRWELL under pions - not observed for the double-stage detectors - is intriguing, and should be clarified (and mitigated) in additional laboratory tests. 

The detector thickness limitation imposed by the SiD experiment calls for the use of extremely thin front-end electronics (a requirement which is, at present, not met by the SRS system). Two alternative readout systems may be suitable for this application: SLAC's KPiX board [24], already beam-tested with THGEM-based detectors [13], and the MICROROC chip developed by the LAL/Omega group and LAPP [11], which was extensively tested with MICROMEGAS detectors; its investigations with THGEM-based detectors is already underway.

Optimization studies on SRWELL detectors (single and double stage), as well as work on larger detectors, are underway. One attractive option is the return to argon-based gas mixtures, implying 2-3 fold higher MIP-induced ionization electrons, though at the cost of higher operation potentials [25]. Modern low-noise electronics may allow for lower-gain operation, so this might not be of a problem.

\section{Acknowledgements}
We are indebted to Hans Muller, Sorin Martoiu, Marcin Byszewski and Konstantinos Karakostas for their assistance with the SRS electronics and associated software. We thank Victor Revivo and Amnon Cohen of Print Electronics Israel for producing the detector electrodes. We thank Rui de Oliveira of CERN for helpful discussions and assistance with the detector electrodes, for producing the readout pad array and for his on-site support during the beam test. We thank Leszek Ropelewski and his team at CERN's Gas Detector Development group, in particular George Glonti, for their kind assistance during the tests. This work was supported in part by the Israel-USA Binational Science Foundation (Grant 2008246), by the Benozyio Foundation and by the FCT Projects PTDC/FIS/113005/2009 and CERN/FP/123614/2011. The research was done within the CERN RD51 collaboration. H. Natal da Luz is supported by FCT grant SFRH/BPD/66737/2009. C. D. R. Azevedo is supported by the SFRH/BPD/79163/2011 grant. A. Breskin is the W.P. Reuther Professor of Research in the Peaceful use of Atomic Energy.

\end{document}